\documentclass[a4paper, 11pt]{article}
\pdfoutput=1 
 
\usepackage{jheppub} 

\usepackage{amsmath}
\usepackage{amssymb}
\usepackage{graphicx,xcolor,subfigure}
\usepackage{empheq}
\usepackage{esint}  
\usepackage{hyperref}
\usepackage[shortlabels]{enumitem}
\usepackage{tikz, pgf} 
\usetikzlibrary{arrows}
\usepackage{float}
\usepackage{slashed}
\usepackage{array}
\usepackage[normalem]{ulem}
\usepackage{tikz}
\usetikzlibrary{shapes.misc}
\tikzset{cross/.style={cross out, draw=black, minimum size=2*(#1-\pgflinewidth), inner sep=0pt, outer sep=0pt},
cross/.default={3pt}}
\usetikzlibrary{decorations.pathmorphing}
\tikzset{snake it/.style={decorate, decoration=snake}}
\usepackage{subfigure,tikz}

\newcommand{\be}{\begin{equation}}
\newcommand{\ee}{\end{equation}}

\title{When the moduli space is an orbifold: Spontaneous breaking of continuous non-invertible symmetries}
\author{Jeremias Aguilera Damia,\ Riccardo Argurio,\ Soumyadeep Chaudhuri}
\affiliation{Physique Th\'{e}orique et Math\'{e}matique and International Solvay Institutes\\
Universit\'{e} Libre de Bruxelles; C.P. 231, 1050 Brussels, Belgium}
\emailAdd{jeremias.aguilera.damia@ulb.be}
\emailAdd{Riccardo.Argurio@ulb.be}
\emailAdd{chaudhurisoumyadeep@gmail.com} 

\abstract{We investigate theories of Nambu-Goldstone bosons where the spontaneously broken continuous symmetry is non-invertible. In such theories, the vacua generically parameterize an orbifold. We study in detail the simplest example of a single free scalar with shift symmetry, modded by reflection symmetry. At singular points of the vacuum manifold, we show that the spectrum of NG excitations is reduced, in particular there are no single-particle states. At the smooth points, on the other hand, single NG modes are present. We show that this is a consequence of the fact that at those points one can construct invertible operators implementing the continuous symmetry on the Hilbert space. }

\begin{document}
\maketitle

\section{Introduction and outlook}
\label{sec:intro}
The spontaneous breaking of a (0-form) continuous global symmetry has profound consequences in quantum field theory (QFT). Most notably, it leads to an effective low-energy description in terms of massless modes, the Nambu-Goldstone (NG) modes, which are directly related to the (Lie algebra) generators of the broken symmetries. Such a low-energy theory can then be formulated as a $\sigma$-model on a group (or more generally a coset) manifold. In this paper we investigate what changes when the continuous symmetry that is spontaneously broken does not form a group, but rather constitutes a fusion category of non-invertible symmetries.

Non-invertible symmetries have been at the center of recent interest, see \cite{Bhardwaj:2017xup, Tachikawa:2017gyf, Chang:2018iay,Thorngren:2019iar,Ji:2019jhk, Rudelius:2020orz, Gaiotto:2020iye, Komargodski:2020mxz, Nguyen:2021yld, Heidenreich:2021xpr, Thorngren:2021yso, Sharpe:2021srf, Koide:2021zxj, Choi:2021kmx, Kaidi:2021xfk, Apruzzi:2021nmk, Roumpedakis:2022aik, Bhardwaj:2022yxj,  Hayashi:2022fkw,Arias-Tamargo:2022nlf, Choi:2022zal, Choi:2022jqy,Cordova:2022ieu, Antinucci:2022eat,Damia:2022seq,Damia:2022bcd,Bashmakov:2022jtl, Niro:2022ctq, Giaccari:2022xgs, Karasik:2022kkq, GarciaEtxebarria:2022jky,Damia:2023ses,Argurio:2023lwl,Cvetic:2023plv} for a partial list of references. While the spontaneous breaking of non-invertible 0-form symmetries of finite order has been considered for instance in \cite{Kaidi:2021xfk,Choi:2022zal,Damia:2023ses,Argurio:2023lwl}, the same question for non-invertible 0-form symmetries of the continuous kind has not yet been explored to our knowledge.\footnote{We stress that our set-up is different from the one in \cite{GarciaEtxebarria:2022jky}, where the broken symmetry is originally of the ``rational" kind considered first in \cite{Choi:2022jqy,Cordova:2022ieu}. The latter is enough to ensure the masslessness of the axion/NG boson. In \cite{GarciaEtxebarria:2022jky} (see also \cite{Karasik:2022kkq}), a construction is devised to see the axion as an ordinary NG boson parameterising an $S^1$, i.e.~the group $U(1)$. This property will be the main distinction from our set-up.}

Continuous non-invertible symmetries are most easily obtained in the following way. Let us start by considering theories with both a global continuous invertible symmetry and a discrete symmetry. Importantly, the latter acts non-trivially on the generators of the continuous symmetry. A paradigmatic example is the one of charge conjugation in presence of a continuous global symmetry.
The non-trivial twist is to then gauge the discrete symmetry acting on the theory. This gauging does not destroy the continuous symmetry, but makes it non-invertible. This is known to happen in  scalar models in two dimensions, {\it e.g.} the $c=1$ orbifold CFT, and in 4$d$ $O(2)$ gauge theory (see for instance \cite{Thorngren:2021yso} and \cite{Heidenreich:2021xpr,Arias-Tamargo:2022nlf,Antinucci:2022eat}, respectively). In both of the cases above, there is no degeneracy of vacua, because in 2$d$ the Coleman theorem prevents the breaking of the symmetry, while in 4$d$ the continuous symmetry is a higher-form symmetry \cite{Gaiotto:2014kfa}.

In order to explore theories with continuous vacuum degeneracies, we generalize the orbifold construction to scalars in $d>2$. In fact, one way to generate such a theory is to put the $O(2)$ gauge theory discussed in \cite{Antinucci:2022eat}, on the manifold $\mathbb{R}^{d-1,1}\times S^1$. In such a set-up the scalar model that we will discuss emerges as a decoupled effective theory in the deep IR through dimensional reduction.\footnote{This is analogous to how a free compactified boson emerges in the deep IR when the Maxwell theory is defined on a manifold with a compact direction.} Our aim is to study the nature of the NG modes arising in this model from the breaking of a continuous non-invertible 0-form symmetry.

In this work we start with a model of a free scalar $\phi$ with a $U(1)$ shift symmetry. This model has a $\mathbb{Z}_2$ reflection symmetry acting as $\phi\to-\phi$ which we proceed to gauge. The $\mathbb{Z}_2$-gauging leads to the shift symmetry becoming non-invertible.\footnote{In fact, the non-invertibility is manifest already in the action on the local charged operators. This is in contrast to many other instances of non-invertible (0-form) symmetries, which usually act invertibly on the local operators charged under them, but manifest their non-invertibility when higher dimensional operators are involved, see e.g.~\cite{Choi:2021kmx,Kaidi:2021xfk,Choi:2022jqy,Cordova:2022ieu}.}
A distinctive feature of this model is that it presents a moduli space of vacua ${\cal M}$ which includes an $S^1/\mathbb{Z}_2$  orbifold. The points in this orbifold are parametrized by a coordinate $\theta\in[0,\pi]$. In ordinary situations, when a modulus originates from the breaking of an invertible symmetry, the Hilbert spaces of NG modes ${\cal H}(\theta)$ are isomorphic at every point in the vacuum manifold. This stems from the fact that there exists a bijective map ${\cal U}: {\cal H}(\theta) \to {\cal H}(\theta')$, generated by the broken symmetry, which is well-defined for every pair of points in ${\cal M}$. Indeed, ${\cal M}$ is a homogeneous manifold in this case. This picture is drastically modified when the symmetry is non-invertible. The vacuum manifold ${\cal M}$ can now have singular points which in our case correspond to $\theta=0$ and $\theta=\pi$. We nevertheless show that there is a non-vanishing order parameter at any point on ${\cal M}$ which means that the non-invertible global symmetry is spontaneously broken everywhere. It turns out however that the Hilbert spaces built upon the singular points are qualitatively distinct from those built upon the other points in the orbifold.

Let us be more specific.
In the presence of global symmetries, we may further refine the definition of the Hilbert space. On general grounds,  $n$-dimensional topological defects associated to a $(d-n-1)$-form global symmetry may host non-trivial $(n-1)$-dimensional (disorder) operators at their boundaries when defined on an open surface. Operators of this kind are usually associated with twisted sectors in the theory. In particular, when $n=1$ and the global symmetry is a $(d-2)$-form $\mathbb{Z}_2$ symmetry, such  operators implement a mapping between two distinct superselection sectors in the Hilbert space of the theory:
\be\label{eq:Hdecomposition}
{\cal H}={\cal H}^{(u)}\oplus {\cal H}^{(t)} \, ,
\ee
where we denote ${\cal H}^{(u)}$ the  untwisted sector and ${\cal H}^{(t)}$ the twisted sector.

In our model, there is a $(d-2)$-form $\mathbb{Z}_2$ quantum symmetry that is generated by topological Wilson lines of the $\mathbb{Z}_2$ gauge field. Hence, the Hilbert space of the theory is split as above.
The spontaneous breaking of the non-invertible 0-form symmetry actually leads to two sets of degenerate vacua in these two superselection sectors. The vacua in the untwisted sector take the form of the $S^1/\mathbb{Z}_2$ orbifold mentioned earlier. However, only the regular points parametrized by $\theta\in(0,\pi)$  have their counterparts in the twisted sector.

Let us emphasize that the moduli space of vacua just described is not affected by the realization of higher form symmetries at low energies when the theory is defined on infinite flat space ${\mathbb R}^{d-1,1}$ \cite{Gaiotto:2014kfa}. The gauged theory actually possesses two types of $(d-2)$-form symmetries. First, there is a continuous non-invertible $(d-2)$-form symmetry acting on (appropriately symmetrized) combinations of integrally quantized vortex operators. Being continuous, this symmetry cannot be spontaneously broken due to a generalized Coleman-Mermin-Wagner theorem. In addition, there is the quantum $(d-2)$-form ${\mathbb Z}_2$ symmetry acting on reflection vortices (more details below). As we will argue in section \ref{discussion}, the fate of this symmetry (i.e. whether it is spontaneously broken or not)  may be different for the regular and the singular points in the vacuum manifold. This, however, does not affect the structure of the moduli space. 

We show that one can define a set of invertible operators acting at any value of $\theta$ lying in the open set $(0,\pi)$. While these operators are not topological, in the sense that there is no sensible (time-like) defect associated to them, they are conserved over time. The action of these operators implements translations between the regular points of the orbifold, while it is trivial on the singular points. Furthermore, these operators define isomorphisms between the Hilbert spaces built upon the different regular points in the orbifold.

Making use of this structure, we establish the existence of single particle states corresponding to the propagating NG modes at any regular point. In fact, one can construct a state with arbitrary number of particles on these vacua. However, at the fixed points of the orbifold, i.e.~$\theta=0$ or $\theta=\pi$, the spectrum is drastically reduced as all odd-particle states are in the twisted sector. In particular, the single particle states are no longer in the untwisted spectrum. This means that the symmetry breaking produces only massless NG bosons in pairs at those specific points.

~

The paper is organized as follows. In section \ref{sec:N=2 case} we discuss in detail our simple model which involves a single compact scalar with shift symmetry and gauged reflection symmetry, i.e.~the simplest orbifold, however in spacetime dimension $d>2$. All the notions that we want to highlight are present in this model: non-invertibility, vacuum degeneracy, vacuum-dependent spectrum. 
In section \ref{discussion} we conclude with some comments on generalizations and further investigations.

\section{\texorpdfstring{$\mathbb{Z}_2$}{Z2}-gauged theory of a free compact scalar}
\label{sec:N=2 case}

\subsection{Review of the free theory of a compact scalar}

Let us consider the theory of a free compact scalar, $\phi\sim\phi+2\pi$ in $d$-dimensional Minkowski  spacetime. The action of this model is given by
\be
\mathcal{S}=\frac12\ g\ \int d\phi\wedge \star d\phi\, ,
\ee
where $g$ is a parameter with mass dimension $d-2$. This model has a $U(1)$ global symmetry realized by shifts $\phi\to \phi-\alpha$ where $\alpha$ is a constant respecting the identification $\alpha\sim\alpha+2\pi$. The associated Noether current 
\be\label{U(1) Noether}
j= -g\, d\phi
\ee
is conserved due to the equation of motion $d\star d\phi=0$. The symmetry is implemented by $(d-1)$-dimensional topological operators
\be\label{U(1) topological}
{\cal U}_\alpha (\Sigma) = e^{i\alpha \mathcal{Q}(\Sigma)}=e^{i\alpha \int_{\Sigma}\star j},
\ee
where $\Sigma$ can be taken to be either a closed submanifold when considering action on operator insertions, or a space-like surface extending to infinity when considering action on the physical states. With a slight abuse of notation we will use the same symbols for both the cases. The meaning should be clear to the reader from the context. Local operators charged under this symmetry are accounted for by properly quantized vertex operators $e^{in\phi(x)}$, $n\in{\mathbb Z}$.

The standard quantization of the free field $\phi$ in momentum space reads
\begin{equation}
\phi(t,\textbf{x})=\overline{\phi}+\lim_{V\rightarrow\infty}\frac{\overline{\pi}}{gV}t+\frac{1}{\sqrt{2g}}\int \frac{d^{d-1}k}{(2\pi)^{d-1}}\frac{1}{\sqrt{|\textbf{k}|}}\Big[a_\textbf{k} e^{-i|\textbf{k}|t+i \textbf{k}.\textbf{x}}+a_\textbf{k}^{\dag}e^{i|\textbf{k}|t-i\textbf{k}.\textbf{x}}\Big]\ ,
\label{N=2: mode expansion of the field}
\end{equation}
with the usual commutation rules, {\it i.e.} 
\begin{equation}
[\overline{\phi},\overline{\pi}]=i\ , \quad \ [a_{\bf k},a_{\bf k'}^\dagger]=(2\pi)^{d-1}\delta^{(d-1)}({\bf k}-{\bf k'})\ .
\end{equation}
Here $\overline{\phi}$ is the zero-momentum mode which satisfies the identification $\overline{\phi}\sim\overline{\phi}+2\pi$. $\overline{\pi}$  is the  momentum conjugate to $\overline{\phi}$. The  term involving  $\overline{\pi}$ vanishes in the infinite volume ($V$) limit. Nevertheless, we indicate it since, even in this limit, $\overline{\pi}$ appears in the charge $\mathcal{Q}$ that generates the shift symmetry given in \eqref{U(1) topological}.\footnote{When the surface $\Sigma$ in \eqref{U(1) topological} is an infinite space-like  surface, $\mathcal{Q}=-\overline{\pi}$.} This shift symmetry  is spontaneously broken resulting in a continuous set of degenerate vacua parametrized by the eigenvalues of  $e^{i\bar\phi}$, namely $e^{i\bar\phi}|\theta\rangle = e^{i\theta}|\theta\rangle$ with $\theta\sim \theta+2\pi$. Therefore, the moduli space takes the form ${\cal M}_0=S^1$. Throughout this work, we will take the spacetime dimension to be $d>2$, so that we are actually dealing with the low energy effective theory of an NG boson. Indeed, in $d=2$ the vacuum degeneracy is lifted due to the Coleman-Mermin-Wagner theorem, while we would like to focus precisely on the properties of the moduli space of vacua.
Shifts within ${\cal M}_0$ are generated by the topological operators \eqref{U(1) topological} acting as\footnote{For sake of notational simplicity, we are omitting the fixed time slice $\Sigma$ in these expressions.}
\be\label{U(1) action}
{\cal U}_\alpha |\theta\rangle = |\theta+\alpha\rangle \quad , \quad {\cal U}_\alpha a_{\bf k} {\cal U}^\dagger_\alpha=a_{\bf k} \quad , \quad {\cal U}_\alpha a_{\bf k}^\dagger {\cal U}^\dagger_\alpha=a_{\bf k}^\dagger\ .
\ee

In addition to the above-mentioned $U(1)$ symmetry, there is also a ${\mathbb Z}_2$ 0-form symmetry which is the reflection $\phi\to-\phi$. Hence, it acts on the conserved current as $j\to-j$. This enhances the symmetry group to 
$U(1)\rtimes \mathbb{Z}_2\cong O(2)$. There is also a $U(1)$ $(d-2)$-form symmetry associated to the topologically conserved current $\hat j=(2\pi)^{-1}\star d\phi$. Objects charged under this symmetry are properly quantized holonomies $e^{in\int\hat\phi}$ of the dual $(d-2)$-form field, defined by $d\hat\phi=2\pi g\star d\phi$, over closed $(d-2)$-dimensional manifolds. Note that the conserved current $\hat j$ is also reversed by the action of the ${\mathbb Z}_2$ reflection symmetry. However, being a continuous $(d-2)$-form symmetry, it can never be spontaneously broken \cite{Gaiotto:2014kfa}. Hence, it does not play any substantial role in our analysis.

Let us notice that the ${\mathbb Z}_2$ reflection symmetry is not realized in the same way for all points in ${\cal M}_0$. In fact, this symmetry is preserved only by the vacua $|0\rangle$ and $|\pi\rangle$, whereas it is broken for the rest of the values of $\theta$. Note that since $O(2)\cong U(1)\rtimes \mathbb{Z}_2$ (and not a direct product), when it is spontaneously broken the moduli space is still just isomorphic to $S^1\cong U(1)$, but with the $\mathbb{Z}_2$  acting non-trivially on all the points except $\theta=0,\pi$.

The Hilbert spaces ${\cal H}_0(\theta)$ associated to the NG bosons are obtained by acting with creation operators $\{a_{\textbf{k}}^\dag\}$ on the respective vacua. The vector spaces ${\cal H}_0(\theta)$ that are obtained in this way from distinct vacua are mutually orthogonal, though isomorphic. The latter statement is a consequence of the fact that different vacua $|\theta\rangle$ and $|\theta'\rangle$ are related by a shift symmetry transformation ${\cal U}_\alpha$ with $\alpha=\theta'-\theta$. From its action \eqref{U(1) action} it is clear that it implements a bijection ${\cal U}_\alpha \, : \, {\cal H}_0(\theta)\to {\cal H}_0(\theta')$. The full Hilbert space of the theory is the direct sum of all these vector spaces.

On general grounds, when a given theory possesses global symmetries, the spectrum of operators decomposes into two classes. On the one hand, the untwisted sector comprises all the {\it genuine} operators in the spectrum. Genuine local operators are the ones that can be defined locally without any need to be attached to topological lines. More generally, an $n$-dimensional operator  is called genuine when it does not live on the boundary of any $(n+1)$-dimensional open topological defect. On the contrary, the twisted sector is formed by all the {\it non-genuine} operators, that is the ones that are well defined only as boundaries of topological defects.\footnote{\label{foot:twisted}This notion becomes more transparent in two spacetime dimensions, where the state operator correspondence induces a similar grading on the Hilbert space. More precisely, non-genuine operators defined at the endpoints of topological lines are in one-to-one correspondence with states obtained by quantization with twisted boundary conditions. This defines the so-called defect Hilbert space. In higher dimensions, this construction becomes less precise, mainly due to the fact that topological defects may come in various dimensionalities. Borrowing the intuition from the two-dimensional case, we will  associate a state in a twisted Hilbert space to operators attached to topological line defects. The latter necessarily correspond to generators of a $(d-2)$-form symmetry. On the contrary, extended non-genuine operators will not be interpreted in terms of states.}

Let us pause here to comment about the different classes of operators arising in this theory. First, genuine local operators are accounted for by properly quantized vertex operators $e^{in\phi(x)}$, $n\in \mathbb{Z}$, together with arbitrary products of derivatives of $\phi(x)$. As explained above, there are also genuine $(d-2)$-dimensional `vortices' described by properly quantized holonomies of the dual field. These exhaust the untwisted sector in the ungauged theory. Let us now list the non-genuine operators contained in the twisted sector.   
Associated to $(d-1)$-dimensional defects of the $U(1)$ 0-form shift symmetry there are twisted sectors encompassing improperly quantized vortices. 
There is also a discrete ${\mathbb Z}_2$ `reflection vortex' living on the boundaries of open ${\mathbb Z}_2$ reflection symmetry defects.
When going around either of these vortex-type operators, the field $\phi$ undergoes $\phi\to\phi-\alpha$ ($\alpha\in[0,2\pi)$) or $\phi\to-\phi$ respectively. In $d>2$ non-compact dimensions, these extended operators do not map to states in a twisted Hilbert space (see footnote \ref{foot:twisted}). The latter actually consists of states created by improperly quantized vertex operators $e^{i\nu\phi(x)}$ with $\nu\notin {\mathbb Z}$, that need to be attached to a topological line associated to the $(d-2)$-form $U(1)$ symmetry with current $\hat j$. These operators may become genuine by gauging discrete subgroups of the $U(1)$ shift symmetry but we will ignore them throughout this paper.

\subsection{Operators and states in the \texorpdfstring{$\mathbb{Z}_2$}{Z2}-gauged theory}

Let us now proceed to the theory obtained by gauging the ${\mathbb Z}_2$ reflection symmetry. Before entering into a detailed discussion of this theory, let us spell out the procedure of  ${\mathbb Z}_2$-gauging to avoid any confusion. One way to implement the $\mathbb{Z}_2$-gauging is to introduce a $U(1)$ gauge field, restrict its holonomies to the $\mathbb{Z}_2$ subgroup via a BF action \cite{Kapustin:2014gua, Gaiotto:2014kfa}, and finally couple the scalar field to this gauge field \cite{Komargodski:2017dmc}. We  follow an equivalent approach where we divide the manifold arbitrarily into simply connected patches. Within each patch the scalar field varies continuously, and the Lagrangian is  given by
 \be
\mathcal{L}=\frac{1}{2}g\partial_\mu\phi\partial^\mu\phi\ .
\label{Lagranigan in a patch}
\ee
However, while going from one patch to another neighboring patch, the field  can undergo a  reflection in the overlapping region. The corresponding transition function is $-1$ or $+1$ depending on whether such a $\mathbb{Z}_2$ transformation takes place or not. A gauge transformation in this picture corresponds to flipping the sign of the field throughout a patch. The path integral involves summing over all field configurations (satisfying the above-mentioned constraints) while identifying configurations that are related by such gauge transformations.\footnote{In such a path integral Dirichlet boundary conditions are imposed at infinity, namely the configurations related by sign flips of the field at infinity are not identified.}

As a consequence of the $\mathbb{Z}_2$-gauging, only the ${\mathbb Z}_2$-neutral sector of the genuine operators discussed above remains genuine. In addition, gauging the ${\mathbb Z}_2$ reflection symmetry retrieves originally non-genuine operators into the spectrum. This occurs with the $(d-2)$-dimensional reflection vortices. Moreover, these are charged under the dual (quantum) $\hat{\mathbb Z}^{(d-2)}_2$ $(d-2)$-form symmetry generated by topological defect lines corresponding to the holonomies of the ${\mathbb Z}_2$ gauge field \cite{Vafa:1989ih,Bhardwaj:2017xup,Tachikawa:2017gyf}.

Let us discuss the spectrum of operators in the $\mathbb{Z}_2$-gauged theory in more detail. The local vertex operators that survive under the gauging are given by the symmetric combinations
\begin{equation}\label{N=2 vertex operator}
\mathcal{V}_n(x)\equiv\ \frac{1}{2}(e^{in\phi(x)}+e^{-in\phi(x)})\quad , \quad n\in {\mathbb Z} \, .
\end{equation}
The antisymmetric combinations, by themselves, are not gauge-invariant. However, one can construct gauge-invariant operators out of them by attaching a semi-infinite topological $\mathbb{Z}_2$ Wilson line as shown below 
\begin{equation}
    \mathcal{W}_n(x) \equiv\ \frac{1}{2i}\left(e^{in\phi(x)}-e^{-in\phi(x)}\right)\eta_{x}^\infty \ ,
    \label{disorder}
\end{equation}
where $\eta_{x}^\infty$ denotes the semi-infinite line ending at the point $x$. $\eta_x^{\infty}$ is given by the product of the transition functions for all the overlapping regions through which the line passes as it goes from one patch to another. The topological nature of the line follows from the flatness of the $\mathbb{Z}_2$-gauge connection. In the simply connected spacetime $\mathbb{R}^{d-1,1}$ that we are considering, all such semi-infinite Wilson lines ending at a particular point are equivalent as there is no loop where the gauge connection has a nontrivial holonomy. The operators in \eqref{disorder}  belong to the spectrum of non-genuine local operators. In fact, since these operators are attached to a line that generates the dual quantum symmetry $\hat{\mathbb{Z}}_2^{(d-2)}$, one may regard them as disorder operators of the latter symmetry.\footnote{As non-genuine operators, they are not unambiguously defined in presence of reflection vortices, i.e.~the objects whose charge is measured by closed loops of the $\mathbb{Z}_2$ gauge field.}

As a consequence of the topological line in \eqref{disorder}, the action  of such an operator can be interpreted as a map from objects in the untwisted sector to those in the twisted sector and vice versa. Importantly, note that the subsector of non-genuine operators is generically not closed under fusion. More precisely, due to the fusion algebra satisfied by the ${\mathbb Z}_2$ topological lines ({\it i.e.} $(\eta_x^{\infty})^2=1$), products of an even number of disorder operators lead to genuine local operators. We will make use of this property in the following analysis.

The above-mentioned semi-infinite Wilson line can also be attached to the field operator $\phi(x)$ yielding the twisted field $\phi^\prime(x)$ defined below:
\begin{equation}
\phi^\prime(x) \equiv\ \phi(x)\eta_{x}^\infty\ .
\end{equation}
The periodicity of the field $\phi(x)$ leads to the identification $\phi^\prime(x)\sim \phi^\prime(x)+2\pi$. 
Let us note that this twisted field satisfies the equation of motion $d\star d\phi^\prime=0$ which follows from the Lagrangian \eqref{Lagranigan in a patch} and the fact that $\eta_x^\infty$ does not vary within a patch.
Moreover, by taking cosines and sines of this field, one can get the operators in \eqref{N=2 vertex operator} and  \eqref{disorder} as shown below:
\begin{equation}
   \mathcal{V}_n(x)=\cos\Big(n\phi^\prime(x)\Big)\ ,\ \mathcal{W}_n(x)=\sin\Big(n\phi^\prime(x)\Big).
   \label{vertex operators in terms of non-genuine field}
\end{equation}
The even powers that appear in the expansion of the cosines lead to the disappearance of the Wilson line since $(\eta_x^{\infty})^2=1$. Similarly, a single factor of $\eta_x^{\infty}$ survives in each term of the expansion of the sines. So, one ends up with the expressions given in \eqref{N=2 vertex operator} and  \eqref{disorder}.

In addition to the twisted field $\phi^\prime(x)$ discussed above, let us introduce its canonical conjugate  
 \be
\pi^\prime(x)\equiv g\partial_t\phi^\prime(x)
\ee
which is also a non-genuine local operator due to the attached Wilson line. Now, we can canonically quantize the fields  $\phi^\prime(x)$ and $\pi^\prime(x)$ and demand the following equal-time commutation relations:
 \be
[\phi^\prime(t,\textbf{x}),\phi^\prime(t,\textbf{y})]=0,\ [\pi^\prime(t,\textbf{x}),\pi^\prime(t,\textbf{y})]=0,\ [\phi^\prime(t,\textbf{x}),\pi^\prime(t,\textbf{y})]=i\delta^{(d-1)}(\textbf{x}-\textbf{y}).
\label{position space commutators between twisted operators}
\ee
Note that the above commutation relations have support only at coincident points. Hence, the effect of the Wilson line trivializes and these commutation relations are the same as those between the field $\phi$ and its conjugate momentum $\pi\equiv g\partial_t\phi$ in any gauge.\footnote{To see this, one simply notes that $\pi^\prime(x)=\pi(x)\eta_x^\infty$, and then uses again the fusion rule $(\eta_x^\infty)^2=1$.}

Next, analogous to \eqref{N=2: mode expansion of the field}, we can do a Fourier mode expansion of $\phi^\prime$ as follows:
\begin{equation}
\phi^\prime(t,\textbf{x})=\overline{\phi}^\prime+\lim_{V\rightarrow\infty}\frac{\overline{\pi}^\prime}{gV}t+\frac{1}{\sqrt{2g}}\int \frac{d^{d-1}k}{(2\pi)^{d-1}}\frac{1}{\sqrt{|\textbf{k}|}}\Big[a_\textbf{k}^\prime e^{-i|\textbf{k}|t+i \textbf{k}.\textbf{x}}+a_\textbf{k}^{\prime\dag}e^{i|\textbf{k}|t-i\textbf{k}.\textbf{x}}\Big],
\label{N=2: mode expansion of the non-genuine field}
\end{equation}
with $\overline{\phi}^\prime\sim \overline{\phi}^\prime+2\pi$. All the Fourier modes are defined by integrals along a spatial slice.They are essentially linear combinations of the twisted operators $\phi^\prime(t,\textbf{x})$ and $\pi^\prime(t,\textbf{x})$ over that slice. Therefore, these Fourier modes, which act on the total Hilbert space of the theory, now map states in the untwisted sector to the twisted sector and vice versa. Note that the total, or extended, Hilbert space includes both untwisted and twisted states, that we will describe in detail shortly.

Based on the commutation relations given in \eqref{position space commutators between twisted operators}, we get the following commutators between the Fourier modes introduced above:
\begin{equation}
[\overline{\phi}^\prime,\overline{\pi}^\prime]=i\ ,\quad \ [a_\textbf{k}^{\prime},a_{\textbf{k}^\prime}^{\prime\dag}]=(2\pi)^{d-1}\delta^{(d-1)}(\textbf{k}-\textbf{k}^\prime).
\label{commutation relations between twisted operators}
\end{equation}
We will shortly use the above Fourier modes and the commutation relations between them to construct the Hilbert space of the theory. For the moment, let us continue our discussion on the operators in the theory.

From the operator $\phi^\prime(x)$ defined above, one can also form an analogue of  the current in \eqref{U(1) Noether} as follows
\begin{equation}
    j^\prime(x) \equiv  -g d\phi^\prime(x) .
    \label{non-genuine current}
\end{equation}
It is tempting to use this non-genuine current operator to construct an analogue of the operator implementing the shift symmetry in the ungauged theory as shown below:
\begin{equation}
    \mathcal{U}^\prime_\alpha(\Sigma)\equiv  e^{i\alpha\int_{\Sigma} \star  j^\prime} .
    \label{combination of genuine and non-genuine operators}
\end{equation}
More precisely, the above construction is needed only for $\alpha\neq0,\pi$ as the operators $\mathcal{U}_0(\Sigma)$ and  $\mathcal{U}_\pi(\Sigma)$ are already gauge invariant by themselves.\footnote{$\mathcal{U}_0$ is just the identity operator, while $\mathcal{U}_\pi$ implements the shift $\phi\rightarrow\phi-\pi\sim\phi+\pi$ which commutes with the $\mathbb{Z}_2$ gauge transformation.}

An operator such as \eqref{combination of genuine and non-genuine operators} is indeed invariant under deformations of the surface $\Sigma$ as can be seen  from the equation of motion $d\star d\phi^\prime=0$. 
However, the operator  $\mathcal{U}^\prime_\alpha(\Sigma)$ is a linear combination of genuine and non-genuine surface operators 
which can be seen  as follows. For simplicity, let us place ourselves in a set up suitable for canonical quantization, i.e.~take $\Sigma$ to be a fixed time slice. Then we have $\mathcal{U}^\prime_\alpha(\Sigma)=e^{-i\alpha \bar\pi'}$, where $\bar\pi'$ is a twisted operator as we saw above. Hence an operator like $\mathcal{U}^\prime_\alpha(\Sigma)$ takes a generic state into a superposition of untwisted and twisted states. Indeed, recalling that the sum of the untwisted and twisted Hilbert spaces is nothing else than the Hilbert space of the ungauged theory, we recognize that $\mathcal{U}^\prime_\alpha(\Sigma)$ acts exactly as $\mathcal{U}_\alpha(\Sigma)$ there. However, in the gauged theory, the operators that truly implement symmetries are only those that act within the untwisted sector (and the twisted sector, separately). It is only the latter that we will call genuine. 

Let us comment here on the following subtlety that needs to be taken into account if one considers such defects on a surface $\Sigma$ with at least a non-trivial one-cycle.
In this case, the definition of the operators \eqref{combination of genuine and non-genuine operators} has an ambiguity due to the fact that while integrating, the semi-infinite line that defines $j'$ can wind (or not) on the non-trivial 1-cycles. 
This is taken care of by positing that the final expression contains as a factor the projector\footnote{ As already mentioned, exceptions to this definition are the operators related to $\alpha=0$ and $\pi$. Those are associated to the subset of invertible symmetries, and as such they should not involve a projector.}
\begin{equation}
    \mathcal{P}(\Sigma)= \frac{1}{|H_1(\Sigma,\mathbb{Z}_2)|} \sum_{\gamma\in H_1(\Sigma,\mathbb{Z}_2)} \eta(\gamma)\ .
\end{equation}
Note that since $\mathcal{P}(\Sigma)\eta(\gamma)=\mathcal{P}(\Sigma)$ for $\gamma\in H_1(\Sigma,\mathbb{Z}_2)$,
a consequence of this fact is that the operators $\mathcal{U}_\alpha'(\Sigma)$ absorb the $\eta(\gamma)$ closed lines, i.e.~the generators of the quantum $\hat{\mathbb{Z}}_2^{(d-2)}$ symmetry (see \cite{Roumpedakis:2022aik} for an extensive exposition on this kind of operators).\footnote{Another consequence is that if one takes such a defect on a surface $\Sigma$ with a cylindrical shape, wrapping a reflection vortex, it will annihilate it, a first hint of non-invertibility.}

One can then extract a genuine topological operator for $\alpha\neq 0,\pi$ by taking the following symmetric combination of operators $\mathcal{U}^\prime_\alpha(\Sigma)$ and $\mathcal{U}^\prime_{-\alpha}(\Sigma)$:
\be\label{N=2 topological operator}
{\cal T}_\alpha (\Sigma) \equiv \mathcal{U}^\prime_\alpha(\Sigma)+ \mathcal{U}^\prime_{-\alpha}(\Sigma)\, .
\ee
The above normalization yields fusion rules of these operators with integer coefficients, as we will see shortly.

Similarly, a non-genuine topological operator can be obtained by taking the anti-symmetric combination of   $\mathcal{U}^\prime_\alpha(\Sigma)$ and $\mathcal{U}^\prime_{-\alpha}(\Sigma)$: 
\be\label{N=2 non genuine topological operator}
{\cal Z}_\alpha (\Sigma) \equiv \mathcal{U}^\prime_\alpha(\Sigma)- \mathcal{U}^\prime_{-\alpha}(\Sigma)\, .
\ee
Such defects turn genuine operators into non-genuine ones, and vice-versa, as we will discuss below. 

A technical remark is in order: the expression \eqref{N=2 topological operator} for the genuine surface operator as a sum of non-genuine  operators $\mathcal{U}^\prime_\alpha$ should be interpreted with some care as, at the end, the operators ${\cal T}_\alpha$ are indecomposable objects. However, introducing $\mathcal{U}^\prime_{\alpha}$ as a formal intermediate step in the construction leads to a more intuitive picture of the underlying structure.

While the operators $\mathcal{U}_0(\Sigma)$ and $\mathcal{U}_\pi(\Sigma)$ form an invertible $\mathbb{Z}_2$ group, the  ${\cal T}_\alpha (\Sigma)$ implement a non-invertible symmetry which is analogous to the non-invertible symmetry in the $O(2)$ gauge theory \cite{Antinucci:2022eat}.  The non-invertibility of this symmetry  is manifest at the level of the fusion rule satisfied by these operators:
\begin{equation}\label{eq:non-inv fusion}
\mathcal{T}_\alpha(\Sigma)\otimes \mathcal{T}_\beta(\Sigma)=\mathcal{T}_{\alpha+\beta}(\Sigma)+\mathcal{T}_{\alpha-\beta}(\Sigma)\ ,
\end{equation}
where we are taking all of $\alpha$, $\beta$, $\alpha+\beta$ and $\alpha-\beta$ to be different than 0 or $\pi$. If $\alpha\pm\beta=0$ or $\pi$, then the right hand side contains an invertible operator, however its coefficient is a projector (or more specifically, and up to a normalization, a condensation defect). For instance:
\begin{equation}\label{eq:non-inv fusion2}
\mathcal{T}_\alpha(\Sigma)\otimes \mathcal{T}_\alpha(\Sigma)=\mathcal{T}_{2\alpha}(\Sigma)+2\mathcal{P}(\Sigma)\mathcal{U}_0(\Sigma)\ .
\end{equation}
The presence of $\mathcal{P}(\Sigma)$ in front of $\mathcal{U}_0(\Sigma)$ is necessary for consistency with the left hand side, and because $\mathcal{U}_0(\Sigma)\equiv \mathbb{I}$ does not carry it, since it is an invertible defect, see a similar discussion in \cite{Antinucci:2022eat, Bhardwaj:2022yxj}.

Within correlation functions, the operators \eqref{N=2 topological operator} and \eqref{N=2 vertex operator} satisfy the following Ward identity
\begin{equation}
\mathcal{T}_\alpha(\Sigma)\mathcal{V}_n(x)=2\cos\Big(n\alpha\ \text{Lk}(\Sigma, x)\Big)\mathcal{V}_n(x)\ ,
\end{equation}
where $\text{Lk}(\Sigma, x)$ denotes the linking number between $\Sigma$ and $x$. For unit linking, one sees that the symmetry operators with $\alpha=\frac{(2k+1)\pi}{2n}$, $k\in \mathbb{Z}$, annihilate $\mathcal{V}_n(x)$, so that they have a non-trivial kernel. This is another manifestation of their non-invertibility. We will later show that this non-invertible symmetry is spontaneously broken in all the vacua of the theory.

\begin{figure}[H]
\centering
\subfigure[]
{
\makebox[.4\textwidth]{
\begin{tikzpicture}
 \draw (-1,0) node[cross] {};
 \draw (-1,0) circle [radius=0.85cm];
 \draw[snake it](-1,0.85)--(-1,2.4);
  \draw (1,0) node[cross] {};
 \draw[snake it](1,0)--(1,2.4);
  \draw (-2.4,0.3) node {$\mathcal{Z}_\alpha(\Sigma)$};
 \draw (-1,-0.32) node {$\mathcal{V}_n(x)$};
 \draw (1,-0.32) node {$\mathcal{W}_n(x)$};
 $\sim$
\end{tikzpicture}
}
}
\hspace{2cm}
\subfigure[]
{
\makebox[.4\textwidth]{
\begin{tikzpicture}
 \draw (-1,0) node[cross] {};
 \draw (-1,0) circle [radius=0.85cm];
 \draw[snake it](-1,0.85)--(-1,0);
  \draw (1,0) node[cross] {};
  \draw (-2.4,0.3) node {$\mathcal{Z}_\alpha(\Sigma)$};
 \draw (-1,-0.32) node {$\mathcal{W}_n(x)$};
 \draw (1,-0.32) node {$\mathcal{V}_n(x)$};
 $\sim$
\end{tikzpicture}
}
}
\caption{Representation of the Ward identities \eqref{eq:non-genuine Ward} satisfied by the non-genuine topological defects and operators. The wiggly lines represent the topological Wilson lines for the ${\mathbb Z}_2$ gauge field, required for gauge invariance.}
\label{fig:ward identities}
\end{figure}
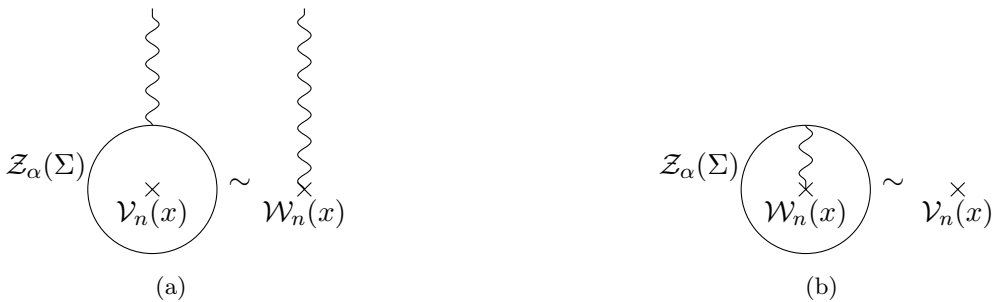

We can similarly display the Ward identities satisfied by the non-genuine topological defects:
\begin{equation}\label{eq:non-genuine Ward}
\begin{split}
&\mathcal{Z}_\alpha(\Sigma)\mathcal{V}_n(x)=2\sin\Big(n\alpha\ \text{Lk}(\Sigma, x)\Big)\mathcal{W}_n(x)\ ,\\
&\mathcal{Z}_\alpha(\Sigma)\mathcal{W}_n(x)=-2\sin\Big(n\alpha\ \text{Lk}(\Sigma, x)\Big)\mathcal{V}_n(x)\ .
\end{split}
\end{equation}
See figure \ref{fig:ward identities} for a diagrammatic representation of these identities.\footnote{Similar diagrammatic representations for topological defects are extensively displayed in the literature, although not necessarily in the same context, see for instance \cite{Chang:2018iay,Bartsch:2023wvv}.}

Finally, an analogous structure arises for the continuous $(d-2)$-form symmetry acting on the vortices since the ${\mathbb Z}_2$ gauge symmetry also takes $\hat j \to -\hat j$. Indeed, one can define (non-)genuine  extended vortices by taking (anti-)symmetric combinations as in \eqref{N=2 vertex operator} and \eqref{disorder}, and similarly for the the continuous $(d-2)$-form symmetry generators.

Having discussed both genuine and non-genuine operators, let us now turn our attention to the states in the theory. We will first discuss the different vacua of the theory. For this, consider the eigenstates of the operator $e^{i\overline{\phi}^\prime}$ which are annihilated by the operators $\{a_{\textbf{k}}^\prime\}$. These are nothing else than the vacua of the ungauged theory, parametrized by the angular variable $\theta$ with $e^{i\overline{\phi}^\prime}|\theta\rangle=e^{i\theta}|\theta\rangle$. Note that just as the operator in \eqref{combination of genuine and non-genuine operators}, $e^{i\overline{\phi}^\prime}$ is a linear combination of genuine and non-genuine operators. Accordingly, its eigenstates are (generically)  linear combinations of states in the untwisted and the twisted sector. To obtain the vacua in these two respective sectors, one needs to take appropriate linear combinations of the above eigenstates.

The vacua in the untwisted sector take the following form:
\begin{equation}
|v^{(u)}\rangle_{\theta}\equiv \begin{cases}\frac{1}{\sqrt{2}}(|\theta\rangle+|-\theta\rangle)\ \text{for}\ \theta\in(0,\pi),\\
|\theta\rangle\hspace{2.3cm} \text{for}\ \theta=0,\pi.\end{cases} 
\label{vacua in untwisted sector}
\end{equation}
The moduli space of vacua in this sector is an orbifold $S_1/\mathbb{Z}_2$ which is parametrized by $\theta\in[0,\pi]$. From these states, one can also obtain the vacua in the twisted sector as follows
\begin{equation}
|v^{(t)}\rangle_{\theta}\equiv \frac{1}{\sin(\theta)}\sin(\overline{\phi}^\prime)|v^{(u)}\rangle_\theta=\frac{1}{\sqrt{2}}(|\theta\rangle-|-\theta\rangle)\ \text{for}\ \theta\in(0,\pi).
\label{vacua in twisted sector}
\end{equation}
They can be thought as parametrizing an open segment.
This can also be inverted to retrieve a subset of the vacua  in the untwisted sector, i.e., 
\begin{equation}
|v^{(u)}\rangle_{\theta}= \frac{1}{\sin(\theta)}\sin(\overline{\phi}^\prime)|v^{(t)}\rangle_\theta
\label{untwisted from twisted}
\end{equation}
In the expressions above, we use $\sin(\overline{\phi}^\prime)/\sin(\theta)$ instead of the simpler expression $\overline{\phi}^\prime/\theta$ because it respects the $2\pi$-periodicity. Note that  $\sin(\overline{\phi}^\prime)$ is a twisted operator.
Furthermore, the action of this operator on the vacua $|v^{(u)}\rangle_0$ and $|v^{(u)}\rangle_\pi$ vanishes. So, there is no counterpart of these vacua in the twisted sector. This illustrates a distinction between these singular points and the regular points in the orbifold. We will later show that this distinction gets carried over to the Hilbert spaces built upon these two classes of vacua.

 Let us make a further comment about our choice of the operator $\sin(\overline{\phi}^\prime)$ to go from the untwisted vacua to the twisted vacua. Instead of the sine function, one could have considered a more general function $f(\theta)$ which is $2\pi$-periodic and odd. We can then immediately see that the corresponding operator $f(\overline{\phi}^\prime)$ has a vanishing action on the vacua $|v^{(u)}\rangle_{0,\pi}$. Moreover, the action of this operator on an untwisted vacuum with $\theta\in(0,\pi)$ gives $f(\overline{\phi}^\prime)|v^{(u)}\rangle_{\theta} = f(\theta) |v^{(t)}\rangle_\theta$. In order to obtain all the vacua in the twisted sector from those in the untwisted sector in this way, we only need the function $f(\theta)$ to be nonzero for any $\theta$ in the open interval $(0,\pi)$. By choosing $f(\theta)=\sin(\theta)$, we just work with the simplest such function.

We may now look at the fate of the non-invertible symmetry given in \eqref{N=2 topological operator} at the different vacua mentioned above. The order parameters for this symmetry are the expectation values of the operators defined in \eqref{N=2 vertex operator}  after a normal-ordering, namely
\begin{equation}
\begin{split}
\ _\theta\langle v^{(u)}|:\mathcal{V}_n(x):|v^{(u)}\rangle_{\theta}&= \cos(n\theta)\ \text{for}\ \theta\in[0,\pi],\\
\end{split}
\end{equation}
where $:():$ indicates the normal-ordering in which the creation operators $\{a_{\textbf{k}}^{\prime\dag}\}$ are pushed to the left of the annihilation operators $\{a_{\textbf{k}}^{\prime}\}$.\footnote{As usual, the normal-ordering removes the divergences  in the expectation values of the composite operators $\mathcal{V}_n(x)$.}
It is clear that there is an infinite number of non-vanishing order parameters at any point in the orbifold. Hence, the non-invertible symmetry is spontaneously broken in all these vacua,\footnote{One can check that the same statement is true for the vacua in the twisted sector.} including those at $\theta=0$ and $\pi$.

Let us next discuss the Hilbert spaces that are constructed upon the different vacua. First, let us focus on the states that can be obtained from the vacua $|v^{(u)}\rangle_0$ and $|v^{(u)}\rangle_\pi$.  One can act with the creation operators $\{a_{\textbf{k}}^{\prime\dag}\}$ on $|v^{(u)}\rangle_{0}$ or $|v^{(u)}\rangle_{\pi}$ to get these states. Since these creation operators are twisted operators, an odd number of them acting on the vacuum leads to a state in the twisted sector. On the other hand, the states in the untwisted sector are obtained by the action of even number of these creation operators on the vacuum. We emphasize that, in particular, there is no single particle excitation in the untwisted sector living upon the vacua $|v^{(u)}\rangle_0$ and $|v^{(u)}\rangle_\pi$. Such excitations rather lie in the twisted sector.

Let us now contrast this with the Hilbert spaces built upon the vacua $|v^{(u)}\rangle_{\theta}$ and $|v^{(t)}\rangle_{\theta}$ with $\theta\in(0,\pi)$. Consider the action of the creation operator $a_{\textbf{k}}^{\prime\dag}$ on the vacuum $|v^{(u)}\rangle_{\theta}$:
\begin{equation}
\begin{split}
\ a_{\textbf{k}}^{\prime\dag}|v^{(u)}\rangle_{\theta}= \frac{1}{\sin
(\theta)}a_{\textbf{k}}^{\prime\dag}\sin(\overline{\phi}^\prime)|v^{(t)}\rangle_\theta.
\end{split}
\end{equation}
The resulting state lies in the twisted sector. Similarly, the action of this operator on $|v^{(t)}\rangle_\theta$ gives a state in the untwisted sector:
\begin{equation}
\begin{split}
\ a_{\textbf{k}}^{\prime\dag}|v^{(t)}\rangle_{\theta}= \frac{1}{\sin(\theta)}a_{\textbf{k}}^{\prime\dag}\sin(\overline{\phi}^\prime)|v^{(u)}\rangle_\theta.
\end{split}
\end{equation}
Based on the above observations, we can define the modified creation/annihilation operators
\begin{equation}
\begin{split}
\widetilde{a}_{\textbf{k}}^{(\theta)}\equiv \frac{1}{\sin(\theta)}a_{\textbf{k}}^{\prime}\sin(\overline{\phi}^\prime),\ \widetilde{a}_{\textbf{k}}^{(\theta)\dag}\equiv \frac{1}{\sin(\theta)}a_{\textbf{k}}^{\prime\dag}\sin(\overline{\phi}^\prime)\ \ \text{for}\ \theta\in (0,\pi).
\end{split}
\end{equation}
The action of these operators retains states in the untwisted/twisted sector in the same sector as they are obtained by taking the product of two operators that map between these sectors. They satisfy the following commutation relations
\begin{equation}
\begin{split}
 [\widetilde{a}_{\textbf{k}}^{(\theta)},\widetilde{a}_{\textbf{k}^\prime}^{(\theta)\dag}]=(2\pi)^{d-1}\delta^{(d-1)}(\textbf{k}-\textbf{k}^\prime) \frac{\sin^2(\overline{\phi}^\prime)}{\sin^2(\theta)}\ .
\end{split}
\end{equation}
Note that these commutation relations reduce to the standard ones in the sectors built upon the vacua $|v^{(u)}\rangle_\theta$ and  $|v^{(t)}\rangle_\theta$. Therefore, one can build a tower of states in the untwisted sector by acting the modified creation operators $\{\widetilde{a}_{\textbf{k}}^{(\theta)\dag}\}$ on the vacuum $|v^{(u)}\rangle_\theta$. A similar tower of states can be built upon the vacuum $|v^{(t)}\rangle_\theta$ in the twisted sector by acting with the same operators. In particular, there are single excitation states in both the untwisted and the twisted sectors living upon the vacua $|v^{(u)}\rangle_\theta$ and $|v^{(t)}\rangle_\theta$ respectively.

\subsection{Translations along the moduli space of vacua}
\label{subsec: local group of topological operators in N=2 case}

In this subsection, we want to show that the regular points of the moduli space of the orbifold model are very similar to points in the moduli space of the ungauged model. Indeed, we can build a charge and invertible operators relating the Hilbert spaces built on any two such points. The non-invertible operators on the other hand generically map to linear combinations of states built from different points. We start with the analysis at regular points, and then proceed to confront with what is obtained with the non-invertible operators.

In the above analysis of the Hilbert space of the $\mathbb{Z}_2$-gauged theory we found that there are two distinct classes of vacua in the untwisted sector. On one hand, there are the ground states $|v^{(u)}\rangle_0$ and $|v^{(u)}\rangle_\pi$, namely the singular points of the $S^1/{\mathbb Z}_2$ orbifold. The Hilbert spaces  in the untwisted sector that are constructed on top of these vacua only contain states with an even number of massless excitations. In particular, there is no single excitation state in this sector. On the other hand, the lowest energy states $|v^{(u)}\rangle_{\theta}$ with $\theta\in(0,\pi)$ can be acted upon by an arbitrary number of the modified gauge invariant creation operators $\{\widetilde a^\dagger_{\bf k}\}$. In other words, the Hilbert spaces living on these  vacua have states with arbitrary number of massless excitations.\footnote{Let us comment here that the single and multi-particle states may be experimentally indistinguishable in a gapless theory. It would be interesting to find a physical setting where the difference in the respective Hilbert spaces is made manifest. 
} 

Notice that the ${\mathbb Z}_2$ gauge symmetry is Higgsed in this second class of vacua as can be verified from the  behavior of the 2-point function of the disorder operator $\mathcal{W}_1$ introduced in \eqref{disorder} at a large  separation between the insertions:
\begin{equation}
\label{disorder_vev}
 \ _\theta \langle v^{(u)}|:\mathcal{W}_1(t,\textbf{x}):\  : \mathcal{W}_1(t,\textbf{y}): |v^{(u)}\rangle_\theta \xrightarrow{|\textbf{x}-\textbf{y}|\rightarrow\infty}\sin^2(\theta) \ .
\end{equation}

We will now argue that the above distinction is emphasized by the presence of a charge operator  $\widetilde{Q}$ in the $\mathbb{Z}_2$-Higgsed sector which generates translations along the moduli space of vacua. The allowed range of translations is constrained to keep the vacuum in the $\mathbb{Z}_2$-Higgsed sector, i.e. these translations do not connect the regular points in the moduli space to the singular points ($|v^{(u)}\rangle_0$ and $|v^{(u)}\rangle_\pi$).  The above feature distinguishes these translations from the familiar case of spontaneous breaking of an ordinary symmetry where the translations along the moduli space cover all the vacua. Despite this distinction, the operators implementing the afore-mentioned translations do form a group which is isomorphic to the group of real numbers under addition.\footnote{This is in contrast to the theory without the $\mathbb{Z}_2$-gauging where, as we mentioned earlier, similar translation operators form a group that is isomorphic to $U(1)$.}  We will show that the presence of the charge $\widetilde{Q}$  generating this group of translations about the regular points in the moduli space leads to isomorphisms between the Hilbert spaces built upon such points. We will also show that, just as in case of ordinary symmetry breaking \cite{Beekman:2019pmi}, the single excitation states built upon these vacua are obtained from plane wave superpositions of local excitations of the charge density. Furthermore, we will demonstrate that this charge density (and hence, the charge operator $\widetilde{Q}$) has a vanishing action on the singular points in the moduli space, viz.  $|v^{(u)}\rangle_0$ and $|v^{(u)}\rangle_\pi$. This is what leads to the absence of the single excitation states in the sectors built upon these vacua.

Let us now proceed to construct the charge $\widetilde{Q}$ that we mentioned above. Consider the non-genuine current operator that was given in \eqref{non-genuine current} and multiply this operator by $\sin(\overline{\phi}^\prime)$ to define a modified current 
\begin{equation}
 \widetilde j(x)\equiv j^\prime(x) \sin(\overline{\phi}^\prime)=  -g d\phi^\prime(x) \sin(\overline{\phi}^\prime).
\end{equation}
Here the choice of the sine function can be motivated by the same considerations as after \eqref{untwisted from twisted}. Just as $j^\prime(x)$, this modified current is also  conserved, i.e. $d\star \widetilde j=0$.
Integrating the Hodge dual of this current over a $(d-1)$-dimensional space-like surface $\Sigma$ that extends to infinity, we can construct the conserved charge operator
\begin{equation}\label{tildeQ}
 \widetilde{Q}\equiv\int_\Sigma \star\widetilde j\ .
\end{equation}
Note that the factor $\sin(\overline{\phi}^\prime)$ in $\widetilde j$ implies a surface integral in order to define the (twisted) zero mode $\overline{\phi}^\prime$. In the following, we will always take this surface to be aligned with $\Sigma$ in \eqref{tildeQ}. Then, considering the Taylor expansion of $\sin(\overline{\phi}^\prime)$, we see that we have a sum of terms each of which involves an even number of integrations, starting with a double integral. One can consequently take the integrands to consist of local insertions pairwise connected by finite Wilson lines. This makes the operator $\widetilde{Q}$ a genuine  operator at the price of fixing  the surface $\Sigma$ to be essentially a spacelike slice.\footnote{In the present case, we can take $\Sigma$ to be a surface with trivial $H_1$, so that we do not need to consider a projector over the closed $\eta$ lines.} In this sense it is not a topological operator, as for instance \eqref{N=2 topological operator}, but merely a conserved operator acting on the Hilbert space that we will use to make the structure of the latter more explicit.  
In other words, the surface operator is still invariant under time translations of the surface. However, there is no clear notion of a defect constructed out of it, hence not complying with the modern view on symmetries.

Due to the presence of the factor $\sin(\overline{\phi}^\prime)$, the operator $\widetilde{Q}$ has a vanishing action on the vacua $|v^{(u)}\rangle_0$ and  $|v^{(u)}\rangle_\pi$. On the other hand, it has a nontrivial action on $|v^{(u)}\rangle_\theta$ for $\theta\in(0,\pi)$ which generates translations along the moduli space. To show this, let us take $\Sigma$ to be a constant time slice and use the mode expansion given in \eqref{N=2: mode expansion of the non-genuine field} to obtain
\begin{equation}
\widetilde{Q}= -g\int d^{d-1}x\ \partial_t\phi^\prime(x) \sin(\overline{\phi}^\prime)= -\overline{\pi}^\prime \sin(\overline{\phi}^\prime)\ .
\label{charge operator: particular expression}
\end{equation}
Using this charge operator as a generator, we can define a set of operators
\begin{equation}
\begin{split}
\widetilde U(\xi)=e^{i\xi\widetilde{Q}},\ \xi\in(-\infty,\infty).
\end{split}
\label{operators generated by the modified charge}
\end{equation}
Note that the charge $\widetilde{Q}$ does not obey any quantization condition, hence the operators are indeed parameterized in ${\mathbb R}$. Consider the action of such an operator with a small value $\epsilon$ of the parameter  $\xi$ on the vacuum $|v^{(u)}\rangle_\theta$ ($\theta\in(0,\pi)$):
\begin{equation}
\begin{split}
\widetilde U(\epsilon)|v^{(u)}\rangle_\theta &=\Big(\mathbb{I}+i\epsilon\widetilde{Q}\Big)|v^{(u)}\rangle_\theta+O(\epsilon^2)=\frac{1}{\sqrt{2}}\Big(|\theta+\epsilon\sin\theta\rangle+|-\theta-\epsilon\sin\theta\rangle\Big)+O(\epsilon^2)\\
&=|v^{(u)}\rangle_{\theta+\epsilon\sin\theta}+O(\epsilon^2).
\end{split}
\label{small translation}
\end{equation}
We see that the action on a vacuum specified by $\theta$ depends on the value of $\theta$ itself. For a finite transformation, we would like to determine the value of $\theta'$ that one obtains by acting with a transformation of (finite) parameter $\xi$ on a vacuum given by $\theta$, i.e.
\begin{align}
    \widetilde U(\xi)|v^{(u)}\rangle_\theta = |v^{(u)}\rangle_{\theta'(\xi;\theta)}\ ,
\end{align}
where we have made explicit that $\theta'$ depends also on the starting point $\theta$.
From above we learn that for small variations of the parameter $\xi$ we have 
\begin{align}
\theta'(\xi+\delta\xi;\theta)=\theta'(\xi;\theta) + \delta\xi \sin\theta'(\xi;\theta)+O(\delta\xi^2) .
\end{align}
 In other words, 
\begin{align}
    \frac{\partial}{\partial\xi}\theta'(\xi;\theta)=\sin\theta'(\xi;\theta)\ 
\end{align}
with $\theta'(0,\theta)=\theta$. Integrating the above differential equation we get
\begin{align}
    \theta'(\xi;\theta) = 2\arctan \Big(e^\xi \tan\frac{\theta}{2}\Big)\ .
\end{align}
Now note that for $\theta\in(0,\pi)$ and for $\xi\in\mathbb{R}$,  $ \theta'(\xi;\theta)\in(0,\pi)$. More precisely, for a fixed $\theta\in(0,\pi)$, $\theta'(\xi;\theta)$ is a monotonically increasing function from $\mathbb{R}$ to $(0,\pi)$. Indeed, one can easily see that for $\xi\to-\infty$, $\theta'\to 0$, while for $\xi\to+\infty$, $\theta'\to\pi$. On the other hand, if $\theta=0$, then $\theta'=0$ for any $\xi$. Similarly if $\theta=\pi$, then $\theta'=\pi$ for any $\xi$. Therefore the operators $\widetilde U(\xi)$ keep the vacua $|v^{(u)}\rangle_0$ and $|v^{(u)}\rangle_\pi$ fixed, while they implement translations between the other vacua.

The fusion rule of the operators $\widetilde U(\xi)$ is straightforward:
\begin{align}
    \widetilde U(\xi) \widetilde U(\eta) = e^{i\xi\widetilde{Q}}e^{i\eta\widetilde{Q}}=e^{i(\xi+\eta)\widetilde{Q}}=\widetilde U(\xi+\eta) \ .
\end{align}
One can indeed be easily convinced that 
\begin{align}
    \theta'(\xi;\theta'(\eta;\theta))=\theta'(\xi+\eta;\theta)\ .
\end{align}
In particular, the action of $\widetilde U(\xi)$ is invertible. It reproduces the additive group of the real numbers. The open segment $\theta\in(0,\pi)$ furnishes a faithful representation, while the endpoints $\theta=0,\pi$ provide trivial representations. Note that the states $|v^{(t)}\rangle_\theta$ are acted upon in exactly the same way as $|v^{(u)}\rangle_\theta$ for $\theta\in(0,\pi)$.

Let us now show that these operators implementing translations between the regular points in the moduli space of vacua also define isomorphisms between the Hilbert spaces built upon those vacua. To see this, consider the states of the following form which span a basis of the Hilbert space living on the vacuum $|v^{(u)}\rangle_\theta$ ($\theta\in(0,\pi)$):
\begin{equation}
\begin{split}
|\Psi^{(\theta)}_{\textbf{k}_1\cdots \textbf{k}_n }\rangle=\widetilde{a}_{\textbf{k}_1}^{(\theta)\dag}\cdots \widetilde{a}_{\textbf{k}_n}^{(\theta)\dag}|v^{(u)}\rangle_\theta.
\end{split}
\end{equation}
If $n$ is odd, we further have
\begin{align}
    |\Psi^{(\theta)}_{\textbf{k}_1\cdots \textbf{k}_n }\rangle={a}_{\textbf{k}_1}^{\prime\dag}\cdots {a}_{\textbf{k}_n}^{\prime\dag}|v^{(t)}\rangle_\theta\ ,
\end{align}
while for $n$ even, we have
\begin{align}
    |\Psi^{(\theta)}_{\textbf{k}_1\cdots \textbf{k}_n }\rangle={a}_{\textbf{k}_1}^{\prime\dag}\cdots {a}_{\textbf{k}_n}^{\prime\dag}|v^{(u)}\rangle_\theta\ .
\end{align}
Since $\widetilde{Q}$, and hence $\widetilde U(\xi)$,  commute with ${a}_{\textbf{k}}^{\prime\dag}$, it immediately follows that\footnote{
Here note that we have $\widetilde U(\xi)\widetilde{a}_{\textbf{k}}^{(\theta)\dag}\widetilde U(-\xi)=\widetilde{a}_{\textbf{k}}^{(\theta')\dag}$ only when acting on $|v^{(u)}\rangle_{\theta'}$. On other vacua, the relation does not come out with unit normalization.}
\begin{equation}
\widetilde{{U}}(\xi)|\Psi^{(\theta)}_{\textbf{k}_1\cdots \textbf{k}_n }\rangle
=\widetilde{a}_{\textbf{k}_1}^{(\theta')\dag}\cdots \widetilde{a}_{\textbf{k}_n}^{(\theta')\dag}|v^{(u)}\rangle_{\theta'}\equiv |\Psi^{(\theta')}_{\textbf{k}_1\cdots \textbf{k}_n }\rangle.
\end{equation}
This evidently defines an isomorphism between the Hilbert spaces built upon $|v^{(u)}\rangle_\theta$ and $|v^{(u)}\rangle_{\theta'}$, for any two $\theta,\theta'\in(0,\pi)$.

Now, let us turn our attention to the charge density that appeared in the integral given in \eqref{charge operator: particular expression}. This charge density is 
\begin{equation}
\widetilde{\rho}(t,\textbf{x})\equiv -g\partial_t\phi^\prime(t,\textbf{x}) \sin(\overline{\phi}^\prime).
\end{equation}
By using the mode expansion of $\phi^\prime$ given in \eqref{N=2: mode expansion of the non-genuine field}, we get the following expression for the Fourier transform of the above charge density at the time $t=0$:
\begin{equation}
\begin{split}
\int d^{d-1}x \ e^{i\textbf{k}\cdot\textbf{x}}\widetilde{\rho}(0,\textbf{x})
&=\widetilde{Q} \delta_{\textbf{k},\textbf{0}}+i \sqrt{\frac{g|\textbf{k}|}{2}}\Big[a_{-\textbf{k}}^\prime -a_\textbf{k}^{\prime\dag} \Big] \sin(\overline{\phi}^\prime).
\end{split}
\end{equation}
 From this we can easily see that the single excitation states living on the vacuum $|v^{(u)}\rangle_{\theta}$ ($\theta\in(0,\pi)$) are given by
 \begin{equation}
\begin{split}
\widetilde{a}_\textbf{k}^{\dag}  |v^{(u)}\rangle_{\theta}=i\sqrt{\frac{2}{g|\textbf{k}|}}\frac{1}{\sin(\theta)}\int d^{d-1}x \ e^{i\textbf{k}\cdot\textbf{x}}\widetilde{\rho}(0,\textbf{x})|v^{(u)}\rangle_{\theta}
\end{split}
\label{single excitations in terms of charge density}
\end{equation}
for\ $\textbf{k}\neq\textbf{0}$. Therefore, as mentioned earlier, these single excitation states are obtained from plane wave superpositions of local excitations of the charge density. Note that, just as the charge $\widetilde{Q}$, the charge density $\widetilde{\rho}(t,\textbf{x})$ has a vanishing action on the singular points in the moduli space of vacua ($|v^{(u)}\rangle_0$ and $|v^{(u)}\rangle_\pi$). This   results in the absence of single excitation states like the ones given in \eqref{single excitations in terms of charge density} on these vacua.

Let us mention here that the above discussion for the $\mathbb{Z}_2$-Higgsed vacua in the untwisted sector goes through for their counterparts in the twisted sector. This means that the operators defined in \eqref{operators generated by the modified charge} also implement  translations in the space of the twisted vacua and define  isomorphisms between the Hilbert spaces living on these vacua. The single excitation states in these Hilbert spaces are  given by expressions analogous to  \eqref{single excitations in terms of charge density}.

Finally, let us discuss the action of the operators $\mathcal{T}_\alpha$ implementing the non-invertible symmetry  on the different vacua. For the following discussion, we take the surface $\Sigma$ in their definition given in  \eqref{N=2 topological operator} to a be a constant time slice. Unlike the operators defined in \eqref{operators generated by the modified charge}, these operators are not unitary. For $\alpha\in(0,\pi)$,  $\mathcal{T}_\alpha$ acting on the different regular points in the moduli space generically produces linear combinations of two vacua as shown below:
\begin{equation}
\mathcal{T}_\alpha|v^{(u)}\rangle_{\theta}
=|v^{(u)}\rangle_{\theta+\alpha}+|v^{(u)}\rangle_{\theta-\alpha}\ ,
\end{equation}
with the understanding that if $\theta+\alpha>\pi$, then $|v^{(u)}\rangle_{\theta+\alpha}\equiv |v^{(u)}\rangle_{2\pi-\theta-\alpha}$, and if $\theta-\alpha<0$, then $|v^{(u)}\rangle_{\theta-\alpha}\equiv |v^{(u)}\rangle_{\alpha-\theta}$. We also have the special cases
\begin{equation}
\begin{split}
&    \mathcal{T}_\theta|v^{(u)}\rangle_{\theta}
=\sqrt{2}|v^{(u)}\rangle_{0}+|v^{(u)}\rangle_{2\theta} \ , \\
&  \mathcal{T}_{\pi-\theta}|v^{(u)}\rangle_{\theta}
=\sqrt{2}|v^{(u)}\rangle_{\pi}+|v^{(u)}\rangle_{2\theta-\pi}
\end{split}
\end{equation}
(with the same understanding as above, and the special case $\mathcal{T}_{\pi/2}|v^{(u)}\rangle_{\pi/2}=\sqrt{2}|v^{(u)}\rangle_{0}+\sqrt{2}|v^{(u)}\rangle_{\pi}$), 
and finally the (invertible) cases (from now on for simplicity we identify $\mathcal{T}_0\equiv \mathcal{U}_0$ and $\mathcal{T}_\pi \equiv \mathcal{U}_\pi$)
\begin{equation}\label{action of T0 and Tpi on vacua}
    \mathcal{T}_0|v^{(u)}\rangle_{\theta}
=|v^{(u)}\rangle_{\theta} , \qquad \mathcal{T}_\pi|v^{(u)}\rangle_{\theta}
=|v^{(u)}\rangle_{\pi-\theta}\ .
\end{equation}
The above expressions define the action of $\mathcal{T}_\alpha$ on $|v^{(u)}\rangle_\theta$ $(\theta\in(0,\pi))$  for all $\alpha\in\mathbb{R}$ because of the  identities $\mathcal{T}_\alpha=\mathcal{T}_{-\alpha}$ and $\mathcal{T}_\alpha=\mathcal{T}_{\alpha+2\pi}$ which can be verified from the definition of these operators given in \eqref{N=2 topological operator}.

From the above action of $\mathcal{T}_\theta$ or $\mathcal{T}_{\pi-\theta}$ on $|v^{(u)}\rangle_{\theta}$, we can see that these operators allow one to make a transition from a regular point in the moduli space to a singular point. This may lead the reader to wonder whether, contrary to our previous analysis, the action of these operators on the single excitation states living on  $|v^{(u)}\rangle_{\theta}$ can produce single excitaton states living on the singular points $|v^{(u)}\rangle_{0,\pi}$. This is indeed not the case as we show below for the action of $\mathcal{T}_\theta$ on a single excitation state built upon $|v^{(u)}\rangle_{\theta}$:
\begin{equation}
\begin{split}
\mathcal{T}_\theta \widetilde{a}_{\textbf{k}}^{(\theta)\dag}|v^{(u)}\rangle_{\theta}
&=\frac{1}{\sqrt{2}}(\mathcal{U}_\theta^\prime+\mathcal{U}_{-\theta}^\prime) a_{\textbf{k}}^{\prime\dag}(|\theta\rangle-|-\theta\rangle)\\
&=\frac{1}{\sqrt{2}} a_{\textbf{k}}^{\prime\dag}(|2\theta\rangle-|0\rangle+|0\rangle-|-2\theta\rangle)=\frac{1}{\sqrt{2}} a_{\textbf{k}}^{\prime\dag}(|2\theta\rangle-|-2\theta\rangle)\\
&=\text{sgn}(\pi-2\theta) \widetilde{a}_{\textbf{k}}^{(2\theta)\dag}|v^{(u)}\rangle_{2\theta} \ ,
\end{split}
\end{equation}
with a similar understanding as above, i.e. when $2\theta>\pi$, then $|v^{(u)}\rangle_{2\theta}\equiv |v^{(u)}\rangle_{2\pi-2\theta}$ and $\widetilde{a}_{\textbf{k}}^{(2\theta)\dag}\equiv \widetilde{a}_{\textbf{k}}^{(2\pi-2\theta)\dag}$.
Note that the terms involving states built upon the singular points neatly cancel in the above expression. A similar argument can be presented for the action of $\mathcal{T}_{\pi-\theta}$ on such a  single excitation state. Note that in particular $\mathcal{T}_{\pi/2} \widetilde{a}_{\textbf{k}}^\dag|v^{(u)}\rangle_{\pi/2}=0$.

Let us now consider the action of $\mathcal{T}_\alpha$ on the singular points $|v^{(u)}\rangle_0$ and  $|v^{(u)}\rangle_\pi$: 
\begin{equation}
\begin{split}
\mathcal{T}_\alpha|v^{(u)}\rangle_{0}
=\sqrt{2}|v^{(u)}\rangle_{\alpha},\ \mathcal{T}_\alpha|v^{(u)}\rangle_{\pi}
=\sqrt{2}|v^{(u)}\rangle_{\pi-\alpha}
\end{split}
\end{equation}
for $\alpha\in(0,\pi)$. We see that these operators acting on the vacua at the singular points in the moduli space  produce the vacua at the regular points (up to a normalization factor). The operator $\mathcal{T}_0$ acts trivially on $|v^{(u)}\rangle_0$ and  $|v^{(u)}\rangle_\pi$, whereas  the operator $\mathcal{T}_\pi$ exchanges these two vacua. Let us note here that the invertible symmetry generated by $\mathcal{T}_\pi$ leads to an isomorphism between the Hilbert spaces built upon $|v^{(u)}\rangle_0$ and $|v^{(u)}\rangle_\pi$.

Unlike the operators $\widetilde U(\xi)$ implementing translations between the regular points in the  moduli space of vacua, the operators $\mathcal{T}_\alpha$ are not generated by a charge. Nevertheless, one can perform an expansion of these operators near $\alpha=0$ as follows:
\begin{equation}
\begin{split}
\mathcal{T}_\alpha=2\mathbb{I}+\alpha^2 \int d^{d-1}x \int d^{d-1}y\ \rho_2(t,\textbf{x},\textbf{y}) +O(\alpha^4)\ ,
\end{split}
\label{expansion of non-invertible symmetry operators}
\end{equation}
where
\begin{equation}
\begin{split}
 \rho_2(t,\textbf{x},\textbf{y})
 &\equiv - g^2\partial_t  \phi^\prime(t,\textbf{x})\partial_t  \phi^\prime(t,\textbf{y}).
\end{split}
\end{equation}
Just as  the single excitation states on the regular points in the moduli space are obtained from the action of the charge density on those vacua, the double excitations on the singular points $|v\rangle_{0,\pi}$ are obtained from $\rho_2(0,\textbf{x},\textbf{y})$ as follows:
\begin{equation}
\begin{split}
  & a_{\textbf{k}_1}^{\prime\dag} a_{\textbf{k}_2}^{\prime\dag}|v^{(u)}\rangle_{0,\pi} =\frac{2}{g}\frac{1}{\sqrt{|\textbf{k}_1||\textbf{k}_2|}}\int d^{d-1}x \int d^{d-1}y\ e^{i\textbf{k}_1.\textbf{x}+\textbf{k}_2.\textbf{y}} :\rho_2(0,\textbf{x},\textbf{y}):|v^{(u)}\rangle_{0,\pi}.
\end{split}
\end{equation}
 Similarly, the other states with even number of excitations can be obtained from the integrands appearing in the higher order terms in the expansion \eqref{expansion of non-invertible symmetry operators}. As a final comment  in this section, let us mention that the states in the twisted sector with odd number of $a_{\textbf{k}}^{\prime\dag}$'s acting on $|v^{(u)}\rangle_{0,\pi}$ can be obtained similarly  from the integrands appearing in an expansion of the twisted operator ${\cal Z}_\alpha$ defined in \eqref{N=2 non genuine topological operator}.

\section{Conclusion and discussion}
\label{discussion}
We conclude with some remarks on generalizations of the model that we considered, and on further investigations. 

So far, we have considered only free field theories, i.e.~theories of NG bosons in the strict IR limit. A natural question is whether the features we have discussed are also present when the spontaneous breaking happens in an interacting theory. We believe the answer is positive, simply because most of the arguments can be phrased in terms of the conserved currents before the gauging of the discrete symmetry.

Let us consider for simplicity the case of a single NG mode, modded by reflection symmetry. Because of the broken shift symmetry, the low energy theory will be organized as a derivative expansion. For instance, the first non-trivial interaction is the quartic higher dimensional operator $(\partial_\mu \phi\partial^\mu\phi)^2$, which is automatically invariant under reflections $\phi\to-\phi$. The current on the other hand, is odd under reflections. The vacuum structure is exactly the same, by definition. Then, using the current, one can build the  operators ${\cal U}_\alpha$ in the ungauged theory, and the  operators ${\cal T}_\alpha$ and $\widetilde{ U}_\alpha$ after gauging.\footnote{In an interacting theory, the expansion \eqref{N=2: mode expansion of the non-genuine field} of the twisted field $\phi'$ is no longer valid. Nevertheless, one can still define a zero momentum mode $\overline{\phi}'$ by taking an average of this field over a spatial slice. Using this one can construct the operator $\sin(\overline{\phi}^\prime)$ which enters in the definition of $\widetilde{ U}_\alpha$.} The single massless excitations on the vacua at the regular points in the orbifold can again be extracted from the action of the charge density associated with $\widetilde{ U}_\alpha$ \cite{Beekman:2019pmi}. The absence of these states in the Hilbert spaces built upon the singular points also follows from the vanishing action of the charge density on the respective vacua.

It would be interesting to explore how our analysis may be extended to theories with multiple scalars. In particular, it would be nice to generalize to cases where the gauged symmetry forms a finite non-abelian group such as $S_N$. A subtely  in this case is that the dual quantum symmetry which arises due to the gauging is non-invertible \cite{Bhardwaj:2022lsg}. This may introduce new complications in defining the twisted operators that played an important role in our analysis. We would like to address these issues in the future.

Let us finally return to our model once more and comment on a complementary way to diagnose whether the vacuum is on a singular or a regular point of the moduli space, i.e.~whether the $\mathbb{Z}_2$ gauge symmetry is preserved or  Higgsed, respectively. Now, we recall the argument that ties the breaking or not of the quantum symmetry to whether the original symmetry was broken or not before gauging \cite{Kapustin:2014gua,Ji:2019ugf,Ji:2019jhk,Su:2023hud}. This is because order parameters for the original symmetry in the ungauged theory become disorder parameters (twisted sectors) for the quantum symmetry upon gauging. Moreover, for a given symmetry, a non-vanishing order parameter implies a vanishing disorder parameter, and vice-versa. In other words, we can probe whether a symmetry is broken or not by the vacuum expectation value of its disorder parameter.

Then, for the quantum symmetry of our orbifold model, the correlator \eqref{disorder_vev} would imply that the ${\mathbb Z}_2$ $(d-2)$-form symmetry is preserved in the vacua $|v^{(u)}\rangle_\theta$ for $\theta$ in $(0,\pi)$. Conversely, at $\theta=0,\pi$ the symmetry might be broken since the VEV of the disorder operator vanishes. Unfortunately, even if these arguments can be safely applied to the study of massive phases, it is not clear whether they hold in presence of gapless excitations.    Therefore, it would be interesting to directly probe the status of the quantum symmetry by evaluating the vacuum expectation value of its order parameter, which is a $(d-2)$-dimensional surface, namely a ``reflection vortex" for $\phi$.
One expects to find an area law in the smooth region of the moduli space, probably after the addition of higher orders in the effective action. In addition, it would be very nice to establish a strong connection between the realization of this emergent symmetry and the structure of the Hilbert space analysed in this work.

An alternative path to reach the same conclusion may be the following. Let us first note that the invertible part of the global symmetry in this theory seems to form an interesting higher group structure.
Indeed, by a simple generalization of the arguments presented for the orbifold theory in $d=2$ \cite{Thorngren:2021yso}, it is easy to check that the ungauged theory contains an anomaly involving all three global symmetries 
\be
S_{anomaly}\supset \pi \int_{d+1} C^{(1)}\cup A^{(1)}\cup B^{(d-1)},
\ee
where $C^{(1)}$, $A^{(1)}$, $B^{(d-1)}$ are the background fields (normalized as integral co-cycles) for the ${\mathbb Z}_2$ reflection symmetry and for the ${\mathbb Z}_2$ restrictions of both the $U(1)$ shift symmetry and the $(d-2)$-form vortex symmetry respectively. We are restricting to the particular subgroups of the continuous symmetries that remain invertible after gauging. 

One can then verify that, upon gauging the reflection symmetry, {\it i.e.} making $C^{(1)}\to c^{(1)}$ dynamical, gauge invariance forces the following correlation between gauge bundles
\be\label{2grpcorrelation}
\delta \hat B^{(d-1)}=A^{(1)}\cup B^{(d-1)} \ ,
\ee
where $\hat B^{(d-1)}$ is the background field  for the quantum symmetry and $\delta$ denotes the co-boundary operator. Indeed, the correlation \eqref{2grpcorrelation} is the signature of a higher group structure \cite{Cordova:2018cvg,Benini:2018reh}.  Such a structure usually leads to interesting hierarchies on the symmetry breaking scales corresponding to the global symmetries involved. In physical terms, such a constraint applied for the case at hand would imply that a phase preserving both ${\mathbb Z}^{(0)}_2$ and ${\mathbb Z}_2^{(d-2)}$ necessarily confines the $\hat{\mathbb Z}_2^{(d-2)}$ reflection symmetry vortices. Now, consider the vacuum at $\theta=\pi/2$. From \eqref{action of T0 and Tpi on vacua}, we see that in this vacuum the ${\mathbb Z}^{(0)}_2$ generated by $\mathcal{T}_\pi$ is unbroken. Furthermore, the generalization of the Coleman theorem states that a continuous $(d-2)$-form symmetry cannot be broken. Thus the $U(1)^{(d-2)}$ vortex symmetry is unbroken in the ungauged model. If we assume that gauging the reflection symmetry does not change that status, then it would follow that its ${\mathbb Z}_2^{(d-2)}$ subgroup that survives in the gauged model is  unbroken as well. If this holds, then we would deduce that by virtue of the higher group constraint, the $\hat{\mathbb Z}_2^{(d-2)}$ reflection symmetry is also unbroken in the $\theta=\pi/2$ vacuum. Recalling the isomorphism between Hilbert spaces discussed in this paper, it would then seem reasonable to extend this conclusion to all regular points of the orbifold. We hope to come back to these problems in the future.

\subsection*{Acknowledgements}
We thank Andrea Antinucci, Giovanni Galati, I\~naki Garcia-Etxebarria, Diego Hofman, Zohar Komargodski, Ho Tat Lam, Giovanni Rizi, Luigi Tizzano, Stathis Vitouladitis and Sasha Zhiboedov for helpful discussions. J.A.D.~and R.A.~are respectively a Postdoctoral Researcher and a Research Director of the F.R.S.-FNRS (Belgium). S.C.~is partially supported by funds from the Solvay Family. The research  of J.A.D., R.A.~and S.C.~is supported by IISN-Belgium (convention 4.4503.15) and through an ARC advanced project.

\bibliographystyle{JHEP} 
\bibliography{Goldstone_noninvertible_references}

\end{document}